\documentstyle[epsfig,aps,prl,multicol]{revtex}
\begin{document}

\title{Electronic polarization in pentacene crystals and thin films}

\author{E.V. Tsiper$^1$ and Z.G. Soos$^2$}

\address{$^1$Department of Chemistry, Rutgers University, 610 Taylor
Rd., Piscataway, NJ 08854}

\address{$^2$Department of Chemistry, Princeton University, Princeton,
NJ 08544}

\maketitle

\begin{multicols}{2}
 [ \begin{abstract}
 Electronic polarization is evaluated in pentacene crystals and in
thin films on a metallic substrate using a self-consistent method for
computing charge redistribution in non-overlapping molecules.  The
optical dielectric constant and its principal axes are reported for a
neutral crystal.  The polarization energies $P_+$ and $P_-$ of a
cation and anion at infinite separation are found for both molecules
in the crystal's unit cell in the bulk, at the surface, and at the
organic-metal interface of a film of $N$ molecular layers.  We find
that a single pentacene layer with herring-bone packing provides a
screening environment approaching the bulk.  The polarization
contribution to the transport gap $P=P_++P_-$, which is 2.01 eV in the
bulk, decreases and increases by only $\sim10$\% at surfaces and
interfaces, respectively.  We also compute the polarization energy of
charge-transfer (CT) states with fixed separation between anion and
cation, and compare to electroabsorption data and to submolecular
calculations.  Electronic polarization of $\sim1$ eV per charge has a
major role for transport in organic molecular systems with limited
overlap.
 \end{abstract} ]

\centerline{PACS: 77.22.-d,71.70.Ch,71.20.Rv}


\section{Introduction}

Prospective molecular electronic devices are based on organic films
deposited on metal or semiconductor surfaces
\cite{horowitz,katz,forrest_review}.  Due to limited mobility, organic
devices are typically restricted to thin (10---100 nm) films.
Variations in electronic polarization energies of charge carriers near
surfaces and interfaces affect the transport states \cite{soos_cpl}
and hence device function.  We have recently addressed electronic
polarization at surfaces and in thin films \cite{ours_surface} using a
self-consistent approach in the limit of vanishing intermolecular
overlap \cite{ours_pplus}, when each molecule experiences the
non-uniform electrostatic potential of all other molecules.  For the
prototypical hole conductor \cite{forrest_review},
perylenetetracarboxylic acid dianhydride (PTCDA), the calculated
transport gap varies by 500 meV between monolayers and thick films, in
agreement with photoelectron and tunneling electron spectroscopy
\cite{ours_surface}.  PTCDA showed that electronic polarization is
accessible to self-consistent computation in crystalline thin films.

In contrast to PTCDA, whose molecules form one-dimensional stacks and
lie almost flat on metallic substrates, many organic molecular
crystals of interest have a herring-bone packing, with molecules
oriented across molecular layers as sketched in Fig.~1.  Both
thiophenes \cite{horowitz} and acenes \cite{katz} have herring-bone
packing and are suitable for thin film transistors.  The molecules in
crystalline thin films are almost upright and charge transport is
preferentially parallel to the surface.

In this paper, we analyze electronic polarization in pentacene,
considering it as a representative of a wider class of organic
materials with electronic applications.  We calculate electronic
polarization in the bulk to obtain the optical dielectric constant,
the transport gap for generating an electron-hole pair at infinite
separation, and the electrostatic binding between molecular ions at
fixed separation in the lattice.  We then consider the experimental
situation in Fig.~1 of thin films on a metallic substrate to compute
electronic polarization at the surface and at the organic-metal
interface, assuming idealized films with structure identical to the
bulk.  Crystalline thin films exhibit multiple phases that, except in
monolayers, are close to the bulk.

 \vskip 0.1 in
 \centerline{\epsfig{file=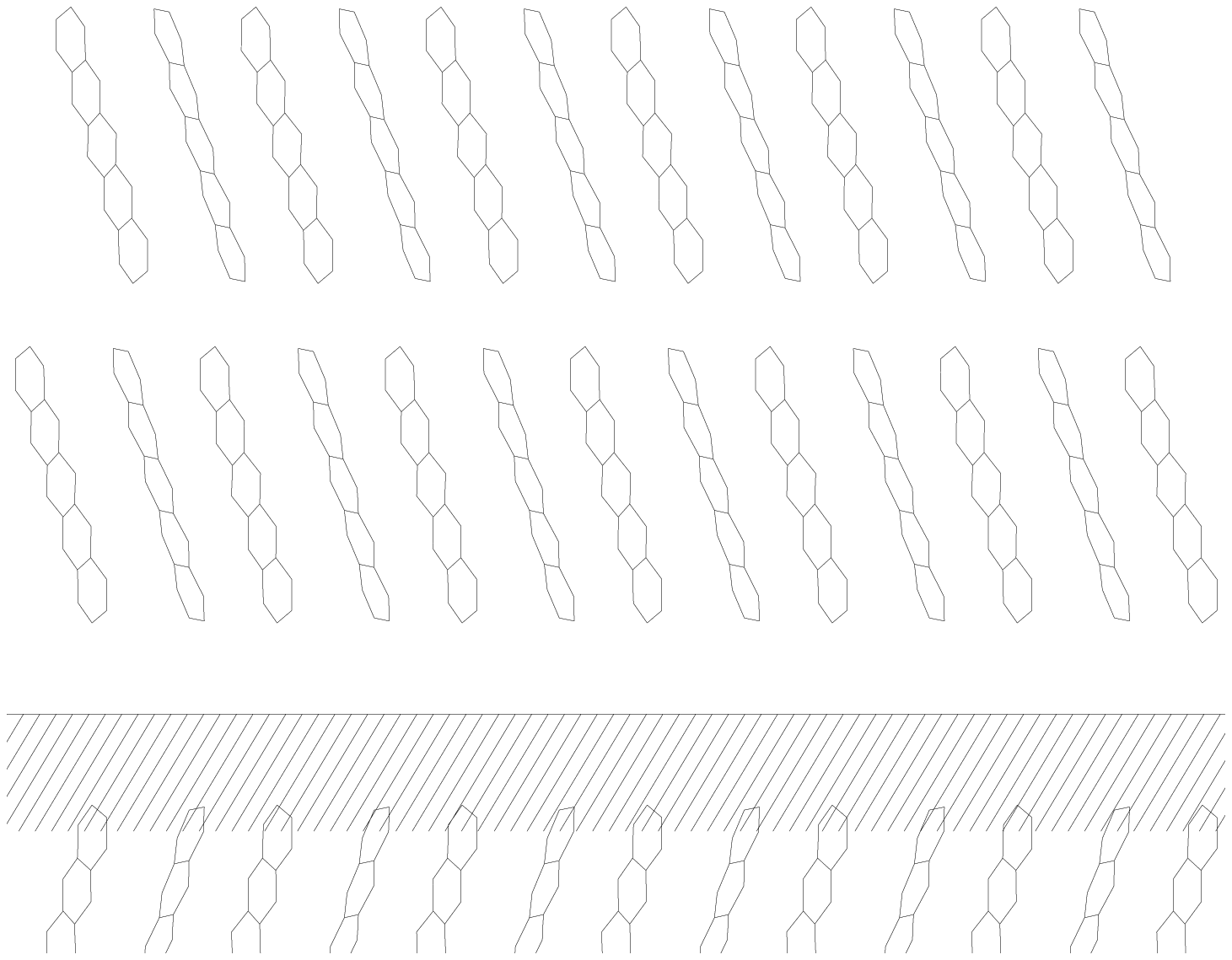,width=3 in}}
 \vskip 0.1 in
 {\small {\bf Fig.~1} Schematic molecular packing in a pentacene layer
on metal.  Image charges below the metal surface are also sketched.}

The long axis of pentacene or $\alpha$-sexithiophene is almost normal to
the surface.  Image charges across the interface in Fig.~1 then
act in the direction of greatest polarizability, which results in
contrasting electronic polarization in pentacene and PTCDA films.
Electric fields normal to the surface induce large dipoles in
pentacenes, but these induced dipoles are parallel and hence
repulsive.  There is competition between charge redistribution due to
image charges and induced dipoles.  Our self-consistent calculations
of charge redistribution indicate that even a single molecular layer
provides a screening environment that reduces repulsion between
induced dipoles.  The improved iterative procedures of Section II are
necessitated by the competition between image charges and induced
dipoles.

The paper is organized as follows.  In Section II we consider the
system of linear equations for electronic polarization and introduce
alternative methods based on a minimum principle when the efficient
iteration scheme of \cite{ours_pplus} fails.  Section III contains
results for bulk crystals, including the optical dielectric constants,
polarization energies of cations and anions, and the binding energies
of ion pairs.  In Section IV we consider pentacene films of 1---10
molecular layers and report polarization energies at the surface and
at the metal-organic interface.  The self-consistent pentacene
potential at the surface illustrates the competition between image
charges and induced dipoles.  Section V contains discussion and
conclusions.

\section{Solution of linear equations}

The zero-overlap approximation reduces electronic polarization to
charge redistribution on molecules in the electrostatic potential of
the solid.  The crystal structure fixes all distances.  We neglect
lattice relaxation whose polarization is estimated to be an order of
magnitude smaller in organic molecular crystals
\cite{silinsh,brovchenko}.  The crystal potential at atoms is readily
introduced as site energies in semiempirical theory, which then
provides a practical approach to computing self-consistent atomic
charges and induced dipoles \cite{ours_pplus}.  We have eight linear
equations per atom and need large ($10^3$) clusters of molecules for
long-range Coulomb interactions.  The large number of equations
precludes the use of algorithms based on transformations of dense
matrices.  For example, a cluster of 2000 PTCDA molecules, as in
\cite{ours_pplus}, leads to 608,000 linear equations for 38 partial
atomic charges and $3\times38$ components of induced atomic dipoles
for each atom of each molecule, and the same number of potentials and
potential gradients.

We review briefly some approaches for solving linear equations that
appear in polarization problems.  We first recast these equations in a
formal but transparent form.  Let $q$ be the vector describing charge
redistribution in a cluster of interest.  Its components may include
partial atomic charges for each atom of each molecule, as well as
induced atomic dipoles.  Higher atomic multipoles can be included in
$q$ as well.  Let $\rho^{(0)}$ be the ``unrelaxed'' charge
distribution, that is the charge distribution in the individual
molecules in the gas phase, the actual charge distribution thus being
$\rho^{(0)}+q$.

For zero overlap, charge conservation for individual molecules leads to
simple constraints on the components of $q$, meaning that $q$ belongs
to a certain subspace ${\cal Q}$ of interest.  We note that
$\rho^{(0)}$ does not necessarily belong to ${\cal Q}$ since the
source charges on ions do not sum to zero.

The polarization problem is defined in terms of the ``state vector''
$q$ as minimization of the total energy $E(q)$, which can be formally
written as

\begin{equation}
E=\frac{1}{2}(qGq) + \frac{1}{2}(\rho^{(0)}+q|V|\rho^{(0)}+q).
\label{E}
\end{equation}
 We use bra- and ket- notation to denote matrix vector products.  The
first term is the energy of non-interacting molecules.  A symmetric
positive-definite {\em stiffness matrix} $G$ ensures that $q=0$ in the
absence of interactions.  It results from solving
Schr\"odinger's equation for a single molecule in non-uniform external
field.  The stiffness matrix is block-diagonal, with separate blocks
corresponding to individual molecules or ions.

The second term describes intermolecular interactions.  The
interaction matrix $V$ contains charge-charge, charge-dipole, and
dipole-dipole interactions, and can also be made symmetric.  The
matrix $V$ is indefinite, reflecting that Coulomb systems are always
unstable.  In contrast to the stiffness matrix, the diagonal blocks of
$V$ are zero, since no molecule interacts with itself.

Expression (\ref{E}) shows a potential for instability, since
it may have no minimum when $V$ is sufficiently large.  In the
organic molecular crystals that we consider, $V$ is small enough and
only results in a shift of the minimum of $E(q)$ from $q=0$.

Differentiating (\ref{E}) with respect to $q$ we find a set of linear
equations on $q$:

\begin{equation}
Gq + Vq = -V\rho^{(0)},
\label{eqGV}
\end{equation}
 subject to the constraint that $q$ belongs to ${\cal Q}$.  While this
form of polarization equations is simple conceptually, another form is
often more practical.  By introducing a set of {\em generalized
potentials} $p=-Gq$ we recast (\ref{E}) in a dual form:

\begin{mathletters}
\begin{eqnarray}
p&=&V(\rho^{(0)}+q)
\label{eqPVa}\\
q&=&-\Pi p,
\label{eqPVb}
\end{eqnarray}
\label{eqPV}
\end{mathletters}
 where $\Pi=G^{-1}$.  These are formally written Eqs.~(8) and (15) of
\cite{ours_pplus}.  The physical meaning of Eq.~(\ref{eqPV}) is
simple.  Instead of a one-step minimization of $E(q)$ as in
(\ref{eqGV}), we optimize the charge distribution of each molecule
individually in the self-consistent external field $p$ of all other
molecules.

The components of the vector $p$ contain potentials and potential
gradients (and possibly higher derivatives), that couple to atomic
partial charges, atomic induced dipoles, and higher atomic multipoles.
The symmetric positive-definite {\em polarization matrix} $\Pi$ has
the same block-diagonal structure as $G$ and describes the
polarization response of individual molecules to the external field,
given by the components of $p$.  The inverse of $G$ is well defined in
the subspace of interest ${\cal Q}$.

\subsection{Iterative solution of polarization equations}

The dual form of Eqs.~(\ref{eqPV}) suggests a natural iterative
approach.  We start with the bare charge distribution $\rho^{(0)}$ and
find ``bare'' potentials $p^{(0)}=V\rho^{(0)}$.  We then repeat
updating charge distribution using Eq.~(\ref{eqPVb}) and recalculating
$p$ using Eq.~(\ref{eqPVa}) until convergence.

When convergent, this procedure is usually fast.  However, it is often
unstable, leading to oscillatory behavior.  A simple numerical trick
has been proven very powerful in suppressing such oscillations.  We
introduce a damping factor $0<f<1$ and iterate, starting with
$q^{(0)}=0$:

\begin{mathletters}
\begin{eqnarray}
p^{(m)}&=&V(\rho^{(0)}+q^{(m)})
\label{itera}\\
q^{(m+1)}&=&fq^{(m)}-(1-f)\Pi p^{(m)}
\label{iterb}
\end{eqnarray}
\label{iter}
\end{mathletters}
 We increase $f$ adaptively when oscillations in $E(q^{(m)})$ are
detected and decrease it when convergence is stable.  All results in
Refs. \cite{ours_pplus,ours_surface,ours_cpl} were obtained using this
iteration procedure, which appears to be both remarkably stable and
fast.  For example, it took just 25 iterations to achieve $10^{-7}$
accuracy in the components of $q$ for the cluster of 2000 molecules of
PTCDA, mentioned above.

Iterations (\ref{iter}) usually perform quite well.  However, they are
not guaranteed to converge and sometimes fail.  For example, we found
intermittent convergence problems for a pentacene cation in a layer on
a metal substrate.  Image charges in the metal induce strong charge
redistribution along the pentacene axis with largest polarizability.
On the other hand, the surrounding molecules in the layer have the
opposite effect of drawing charge into the inside of the layer, closer
to the center of the ion.  As follows from the self-consistent
solution, the second effect outperforms the first.  This clearly leads
to difficulties because the initial iterations tend to drive the
system away from the convergence point, indicating also that a
perturbative solution may be counterintuitive in this case.

Computational difficulties of this kind are rare.  We now describe an
alternative approach to solving polarization equations, which is
formally stable and is guaranteed to converge, while it may not be as
efficient as Eqs.~(\ref{iter}).

\subsection{Subspace methods for polarization problems}

Let us return to the original Eq.~(\ref{eqGV}).  We can treat it
as a general linear system

\begin{equation}
Aq=b,
\label{linear}
\end{equation}
 with a symmetric matrix $A=G+V$ on the left-hand side.  Solution of
such a linear system is equivalent to minimizing the energy functional

\begin{equation}
E(q)=\frac{1}{2}(qAq)-(bq),
\label{energy}
\end{equation}
 whose minimum exists when $A$ is positive-definite.

We present a variationally-stable subspace approach to solving
(\ref{linear}) based on the minimum principle (\ref{energy}).  By
re-scaling the solution we can assume the right-hand side to be
normalized, $(bb)=1$.  We shall be searching for the solution $q$ in
the Krylov subspace ${\cal K}$ of the matrix $A$, generated by the
vector $b$: ${\cal K}=\text{span}(b,Ab,A^2b,...,A^{m-1}b)$.  At every
step $m$ we add a new vector to the subspace ${\cal K}$
and construct the best vector $q$ which minimizes (\ref{energy})
within ${\cal K}$.  This guarantees that each step $m$ yields a better
solution, and thus the procedure converges.

We use the Hermitian Lanczos recursion \cite{lanczos,parlett} to build
an orthonormal basis in ${\cal K}$: $q^{(1)}=b$,

\begin{equation}
q^{(m+1)}=\beta_{m+1}^{-1}(Aq^{(m)}-\alpha_mq^{(m)}-\beta_mq^{(m-1)})
\label{lanc}
\end{equation}
 where $\alpha_m$ and $\beta_m$ are chosen at each step $m$ to
orthogonalize $q^{(m+1)}$ with respect to both $q^{(m)}$ and
$q^{(m-1)}$, ensuring that $(q^{(m)}|q^{(n)})=\delta_{mn}$ for every
$m$ and $n$.  Expressing $q=\sum_mc_mq^{(m)}$ and solving for the
coefficients $c_m$ that minimize $E(q)$ we find that they obey a
linear system $\sum_n\widetilde A_{mn}c_n=\widetilde b_m$, where
$\widetilde A_{mn}=(q^{(m)}Aq^{(n)})$ and $\widetilde b_m=(q^{(m)}b)$.

The algebraic properties of Lanczos recursion (\ref{lanc}) yield
readily all the components of $\widetilde A$ and $\widetilde b$.  The
matrix $\widetilde A$ is symmetric tridiagonal with the diagonal
elements $\widetilde A_{mm}=\alpha_m$ and sub-diagonal elements
$\widetilde A_{m,m-1}=\beta_m$.  Due to the special choice of
$q^{(1)}=b$, all $\widetilde b_m=0$, except $\widetilde b_1=1$ and
$\widetilde b_2=\alpha_1$.

The method described above is variationally stable, since it is based
on a minimum principle.  Its drawback, however, is that it is
generally slower than iterations (\ref{iter}).  The reason is that
Eqs.~(\ref{iter}) follow from the dual form of Eqs.~(\ref{eqPV}),
which is based in turn on a physical insight of solving each molecule
separately in the external field of other molecules.  The subspace
approach is not based on such an insight.  In practice, it takes
200---300 iterations, rather then 25-35, to achieve convergence with
this approach.  Thus we use the subspace approach only when iterations
fail.

\subsection{Subspace method for non-positive definite cases}

The subspace method of the above section being applied to
Eq.~(\ref{eqGV}) deals with the stiffness matrix $G$.  It is
preferable numerically to work with $\Pi$, rather than with its
inverse.  The reason is that whenever there is a small polarizability,
e.g. in the direction normal to the plane of a $\pi$-conjugated
molecule, it translates into a small eigenvalue of $\Pi$ and,
correspondingly, into a large eigenvalue of $G$.  Numerically, it is
much easier to handle near-zero than near-infinity.

In a more drastic situation, $G$ may have negative eigenvalues, yet
the polarization self-consistent equations (\ref{eqPV}) continue to
make sense.  For example, we have introduced atomic polarizabilities
$\widetilde\alpha$ \cite{ours_pplus} as the difference
$\alpha-\alpha^{\rm C}$ between the best available molecular $\alpha$
and $\alpha^{\rm C}$ based on semiempirical atomic charges.  Usually
$\alpha^{\rm C}$ underestimates $\alpha$ by $\sim10$\%.  But the
charge-induced polarizability occasionally exceeds $\alpha$ in some
directions (see e.g. Table I), and the correction $\widetilde\alpha$
has one or more negative eigenvalues.  The self-consistent
polarization equations (\ref{eqPV}) then give a saddle point of the
energy functional, which formally has no minimum.

In order to use the subspace method to deal with $\Pi$ rather than
$G$, we rewrite Eqs.~(\ref{eqPV}) in a matrix form in such a way that
the matrix on the left-hand side is symmetric:

\begin{equation}
\left(\begin{array}{cc}
-V & 1 \\
1 & \Pi 
\end{array}\right)
\left(\begin{array}{c}
q \\
p
\end{array}\right)
=
\left(\begin{array}{c}
p^{(0)} \\
0  
\end{array}\right).
\label{eqM}
\end{equation}
 This is again a linear system of type (\ref{linear}), but of twice
the size.  The matrix is, however, not positive-definite, and so the
above method based on the minimization of (\ref{energy}) cannot be
applied.

We present a modified subspace approach that is applicable to linear
systems of type (\ref{linear}) with symmetric matrix $A$ that is not
necessarily positive definite.  Instead of searching for the minimum
of (\ref{energy}) we minimize the residual norm.

\begin{equation}
Z(q)=\frac{1}{2}(Aq-b|Aq-b),
\label{residual}
\end{equation}
 which is also equal to half the norm of the gradient, $\partial
E(q)/\partial q=Aq-b$.  Again, expressing the solution as
$q=\sum_mc_mq^{(m)}$, but solving now for $c_m$ that minimize $Z(q)$
we find another linear system, $\sum_n\widetilde Z_{mn}c_n=\widetilde
w_m$, where $\widetilde Z_{mn}=(q^{(m)}A^2q^{(n)})$ and $\widetilde
w_m=(q^{(m)}Ab)$.

After some algebra, we find that the matrix $\widetilde Z$ is
symmetric 5-diagonal with the following nonzero elements: $\widetilde
Z_{mm}=\alpha_m^2+\beta_m^2+\beta_{m-1}^2$, $\widetilde
Z_{m,m-1}=(\alpha_m+\alpha_{m-1})\beta_{m-1}$, and $\widetilde
Z_{m,m-2}=\beta_{m-1}\beta_{m-2}$, while the right-hand side has all
components zero except $\widetilde w_1=\alpha_1$ and $\widetilde
w_2=\beta_1$.  The above expressions for the elements of $\widetilde
Z_{mn}$ can be found by consequently multiplying Eq.~(\ref{lanc}) by
itself with $m$ replaced with $m-1$, $m-2$, etc.  We notice by
inspection that $\widetilde Z=\widetilde A^2$, except for the last row
and column.

Throughout this section we have assumed that all vectors are projected
onto the subspace ${\cal Q}$ to preserve charge conservation
relations.  We did not discuss preconditioning techniques, such as
rescaling the variables, which may improve convergence.

\section{Polarization in Pentacene Crystals}

\subsection{Gas-phase properties}

We performed calculations of the electronic structure of pentacene in
gas phase using B3LYP hybrid density functional theory with the
extensive 6-311++G(d,p) basis set available in Gaussian 98
\cite{gaussian}.  D2h symmetry was assumed and the geometry was
separately optimized for the neutral molecule, cation, and anion.
Unrestricted Hartree-Fock (UHF) was used for the ions.  Limited spin
contamination was observed as estimated by the maximum deviation of
the total spin $\overline{S^2}=0.769$ before annihilation.  Comparison
of the energies of the ions in the neutral and ionic optimized
geometries yields the relaxation energies $\lambda_+=46$ meV for the
cation and $\lambda_-=68$ meV for the anion, under the assumption of
equal zero-point contributions for the neutral molecule and ions.
$\lambda_+$ compares well with the value of 59 meV derived from
experiment \cite{kahn_bredas_2002}.

The vertical ionization potential is found to be $I=6.229$ eV, which
is significantly below the recently reported experimental value of
6.589 eV \cite{kahn_bredas_2002}.  Other values of 6.64 eV
\cite{clark} and 6.74 eV \cite{boschi} were reported previously.  The
calculated value of $I$ must be corrected by the difference in
zero-point energies before the direct comparison can be made, but
zero-point differences are probably less than 360 meV.

The experimental value for the electronic affinity is known with an
error bar.  Our calculated B3LYP value $A=1.475$ eV is 105 meV above
the recommended average 1.37 eV \cite{sato} of experimental data.
Thus, B3LYP values lead to the gas-phase pentacene charge gap
$I-A=4.754$ eV (without zero-point correction) vs. 5.22 eV derived
from experiment.

\subsection{Polarization energies}

Following the procedure in \cite{ours_pplus} we first calculate
atom-atom polarizabilities $\Pi_{ij}$ for individual pentacene
molecules using the semiempirical INDO/S Hamiltonian \cite{indos}.
Experimental geometries from the crystal structure data
\cite{pentacene_structure} were used.  Positions of hydrogen atoms,
not known experimentally, were AM1-optimized using Gaussian 98 with
the heavy atoms fixed.

Due to the inequivalence in crystal field for the two molecules in the
unit cell, their geometries are slightly different.  There are two
atom-atom polarizability tensors $\Pi_{ij}$ with $i$, $j$ ranging over
36 atoms of a pentacene,

\begin{equation}
\Pi_{ij}=-\frac{\partial\rho_i}{\partial\phi_j}
   =-\frac{\partial^2E}{\partial\phi_i\partial\phi_j}
\label{Pi}
\end{equation}
 $E$ is the INDO/S ground-state energy, $\phi_i=\phi({\bf r}_i)$ is
the crystal potential at atom $i$ and $\rho_j$ are L\"owdin atomic
charges.  Charge redistribution is then described explicitly as

\begin{mathletters}
\begin{eqnarray}
\rho_i&=&\rho_i^{(0)}-\sum_j\Pi_{ij}\phi_j
\label{SC1a}\\
\bbox{\mu}_i&=&\widetilde\alpha_i\bbox{F}_i,
\label{SC1b}
\end{eqnarray}
\label{SC}
\end{mathletters}
 [cf. Eqs~(\ref{eqPV})], where $\rho_i^{(0)}$ are gas-phase charges,
$\bbox{\mu}_i$ are induced atomic dipoles and
$\bbox{F}_i=-\nabla\phi({\bf r}_i)$ are electric fields.
$\widetilde\alpha=\alpha-\alpha^{\rm C}$ is a correction
\cite{ours_pplus} that accounts for the difference in {\em ab-initio}
molecular polarizability and the ``charges-only'' part $\alpha^{\rm
C}$ associated with $\Pi_{ij}$:
 
\begin{equation}
\alpha^{\rm C}_{\alpha\beta}=\sum_{ij}\Pi_{ij}r_i^\alpha r_j^\beta.
\label{alpha}
\end{equation}
 We computed $\alpha$ for the two geometries using B3LYP density
functional with 6-311++G(d,p) basis set of Gaussian 98.

Principal values of $\alpha$ and ``charges-only'' $\alpha^{\rm C}$ are
compared in Table I.  Charge redistribution accounts quite well for
the in-plane components $\alpha_{\rm LL}$ and $\alpha_{\rm MM}$ of the
neutral molecule.  The polarizability $\alpha_{\rm NN}$, normal to the
molecular plane, is completely ``atomic,'' as expected, and cannot be
described by charge redistribution \cite{ours_cpl,ours_pplus}.  Values
that are not exactly zero are due to non-planar molecules in the
crystal. The INDO/S results for ions are far too large; we return to
this point in the Discussion.

 \vskip 0.1 in
 {\small {\bf Table I.} Molecular polarizability of pentacene along
principal axes ($L=\text{long}$, $M=\text{medium}$,
$N=\text{normal}$)}

 \centerline{
\begin{tabular}{lccc}
\\
\tableline
\tableline
Molecule &
$\alpha_{\rm NN}$ (\AA$^3$) &
$\alpha_{\rm MM}$ (\AA$^3$) &
$\alpha_{\rm LL}$ (\AA$^3$)
\\
\tableline
\\
\multicolumn{2}{l}{\bf \ \ \ \ B3LYP/6-311++G**} & & \\
Neutral(1) &  17.77 & 37.66 &  91.43 \\
Neutral(2) &  18.02 & 38.02 &  99.48 \\
Anion(1)   &  19.87 & 40.16 & 135.47 \\
Anion(2)   &  20.20 & 40.81 & 131.88 \\
Cation(1)  &  16.31 & 35.91 & 130.87 \\
Cation(2)  &  16.54 & 36.54 & 124.14 \\
\\
\multicolumn{2}{l}{\bf \ \ \ \ $\alpha^{\rm C}$, Eq.~(12)} & & \\
Neutral(1) &  0.06  & 39.26 &  87.41 \\
Neutral(2) &  0.05  & 39.61 &  95.83 \\
Anion(1)   &  0.06  & 35.26 & 194.35 \\
Anion(2)   &  0.05  & 39.98 & 231.24 \\
Cation(1)  &  0.06  & 36.03 & 368.28 \\
Cation(2)  &  0.05  & 40.55 & 282.52 \\
\\
\tableline
\\
\end{tabular}}

The difference between the B3LYP and $\alpha^{\rm C}$ in Table I is
distributed over the atoms with the weight proportional to the atomic
valence charge \cite{ours_pplus}.  Since charge redistribution
overestimates the in-plane molecular polarizability in pentacene, the
atomic correction tensor $\widetilde\alpha=\alpha-\alpha^{\rm C}$ is
not positive definite.  It is still small, however, for neutral
molecules, and the self-consistent solution is well defined, even though
it does not correspond strictly to the minimum of the total energy
(cf. discussion in II.C).

\subsection{Properties of the neutral lattice}

Both pentacene \cite{pentacene_structure} and anthracene
\cite{anthracene_structure} illustrate herring-bone packing with two
molecules per unit cell, but their Bravais lattices are different.
Pentacene is triclinic \cite{pentacene_structure}.  The two molecules
are inequivalent and subject to different crystal-field environments.
The polarization energy of a cation or anion consequently depends on
which molecular site is charged.  Since crystalline electric fields
are perturbations, we expect comparable polarization energies that
differ at most by $\sim100$ meV.  Such differences have already been
computed \cite{petelenz_slawik_1996} in pentacene crystals using the
submolecular method \cite{silinsh,munn} in which the gas-phase
molecular polarizability $\alpha$ is placed as $\alpha/5$ at the ring
centers.

Solving self-consistent polarization equations for the neutral lattice
of pentacene, we find that polarization contribution to sublimation
energy is negligible (1.23 meV per molecule), as in anthracene, since
the gas-phase charges are small due to an approximate electron-hole
symmetry of the valence shell.  We find the dielectric tensor of pentacene
to be highly anisotropic, with principal values $\kappa_1=5.336$,
$\kappa_2=3.211$, and $\kappa_3=2.413$.  For reference purposes, the
directional cosines of the principal axes are ${\bf
n}_1=(-0.296,-0.314,0.902)$, ${\bf n}_2=(-0.021,0.946,0.322)$, and
${\bf n}_3=(0.955,-0.076,0.286)$, respectively, in the Cartesian
coordinate system with the pentacene lattice vectors (in\ \AA) ${\bf
a}=(7.900,0,0)$, ${\bf b}=(0.444,6.044,0)$, and ${\bf
c}=(-6.153,-2.858,14.502)$.  The direction of ${\bf n}_1$ coincides
with the direction of long axes of pentacene molecules to the accuracy
to which the latter can be defined ($\sim5^o$).

The calculated principal axes and values of $\kappa$ for anthracene
crystals agree with experiment \cite{ours_cpl}.  Measurements of
refractive indices or optical dielectric constants are challenging in
anisotropic molecular crystals \cite{karl_epsilon_anthracene}.  We are
not aware of such data for pentacene crystals.  Dielectric data will
be necessary for organic devices with improved performance.

\subsection{Charge carriers in the bulk}

We compute the polarization energy of charge carriers following
\cite{ours_pplus}, by placing an ion in the infinite neutral lattice
and considering an imaginary sphere of $M$ molecules surrounding it.
We allow only the molecules whose centers fall into the sphere to
relax their charge distributions from the self-consistent values for
the neutral lattice and monitor convergence as $M$ is increased.  We
use the same molecular geometry for the ion.  The small ($\sim100$
meV) relaxation energies of large aromatic molecules make the same
geometry both an excellent and convenient approximation.  Atom-atom
polarizabilities $\Pi_{ij}$ for the ion are computed using
unrestricted Hartree-Fock (UHF) INDO/S.

 \vskip 0.1 in
 \centerline{\epsfig{file=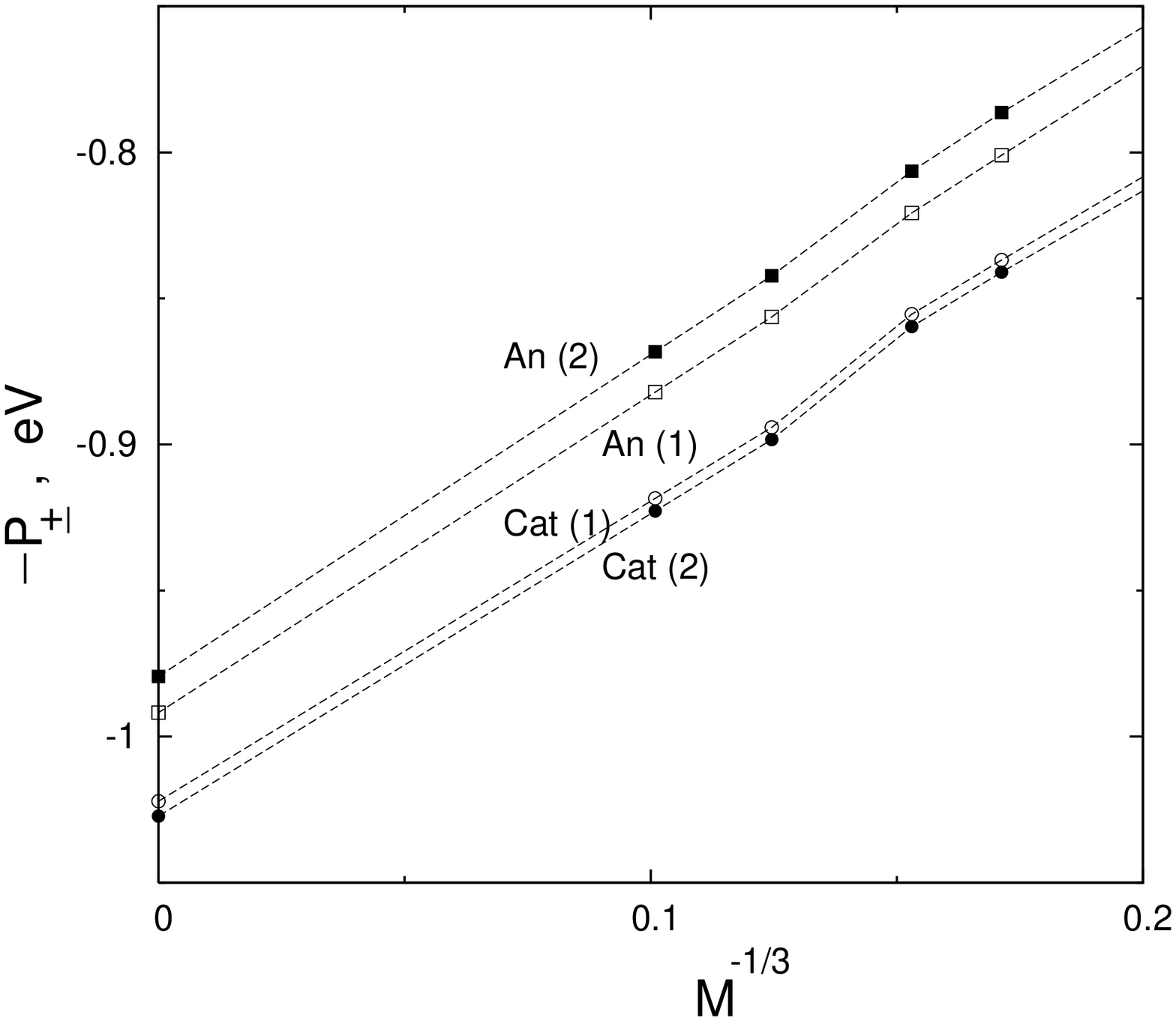,width=2.5 in,angle=-90}}
 \vskip 0.1 in
 {\small {\bf Fig.~2} Convergence of $P_\pm$ with increasing cluster
size $M$.}
 \vskip 0.1 in

Figure 2 shows linear convergence of polarization energy with
$M^{-1/3}$, which is proportional to the inverse radius of the sphere.
The polarization energy extrapolates as $\propto M^{-1/3}$, because
the missing part at large finite $M$ can be thought of as polarization
of an infinite homogeneous dielectric with a spherical cavity of
radius $R\propto M^{-1/3}$, $-(1-\kappa_{\rm eff}^{-1})e^2/R$
\cite{ours_pplus}.  Extrapolation to infinite radius gives the
polarization energy in the bulk, $P=P_++P_-=2.014$ eV and $2.007$ eV
for the ion occupying positions 1 and 2 in the unit cell,
respectively.  The ``charges only'' values of $P=P_++P_-$ are
$\sim10$\% less, as also found in other organic crystals.  Since
$\kappa_{\rm eff}$ for an anisotropic dielectric medium is known
\cite{bounds}, the slope in Fig.~2 at large $M$ is directly related to
the $\kappa_\alpha$ found above, and the slopes agree within a few
percent.  We note that $\kappa_{\rm eff}^{-1}$ implies identical
slopes at large $M$ for the anion and cation in either position, as
found separately in Fig.~2.

Using B3LYP values for $I$ and $A$ in the gas phase, we calculate the
transport gap of the pentacene crystal $E_t=I-A-P=2.740$ eV in the
limit of zero overlap.  This correlates quite well with the reported
band gap of 2.85 eV, obtained from careful fit of electroabsorption
spectra \cite{petelenz_slawik_1996}.  $E_t=2.78$ eV was reported in
\cite{bounds_1985} also based on electroabsorption spectra of
charge-transfer (CT) states.  The experimental value for the charge
gap, $I-A=5.22$ eV, used in this context results in $E_t=3.21$ eV,
which is too high.

\subsection{Ion pairs and CT states}

We next report polarization energies $P_{\rm pair}$ of CT states with
an ion at the origin and the counter ion at a nearby site.  We place
the ion pair within an imaginary sphere in a neutral lattice and
calculate the effective interaction $V_{\rm eff}$, which is the
difference of the total polarization energy and the energy of a pair
of well-separated charges, $V_{\rm eff}=(P_{\rm pair}-P_+-P_-)$.
$-V_{\rm eff}$ is the binding energy of CT states in the limit of no
overlap and is closely related to the binding energies of Frenkel
excitons, when the final state is a molecular excitation.

Fig.~3 shows convergence with the size of the sphere.  Since the total
ion charge within the sphere is zero, the leading term $\propto
M^{-1/3}$ vanishes.  The polarization energy of a dipole in a
dielectric medium converges much faster as $\propto M^{-1}$.
Extrapolating to infinite $M$, we obtain the lowest CT states at
$-0.719$ and $-0.679$ eV for the nearest-neighbors (center-to-center
distance 4.799\ \AA) and $-0.685$ and $-0.675$ eV for next-nearest
neighbors (5.151\ \AA), as listed in Table II.  These results compare
reasonably with submolecular calculations \cite{petelenz_slawik_1996}
of $-0.777$ and $-0.698$ eV, respectively, with $\alpha/5$ at ring
centers.  Submolecular results do not distinguish between anions and
cations and also depend on the precise partitioning of $\alpha$, which
is left open.

 \vskip 0.1 in
 \centerline{\epsfig{file=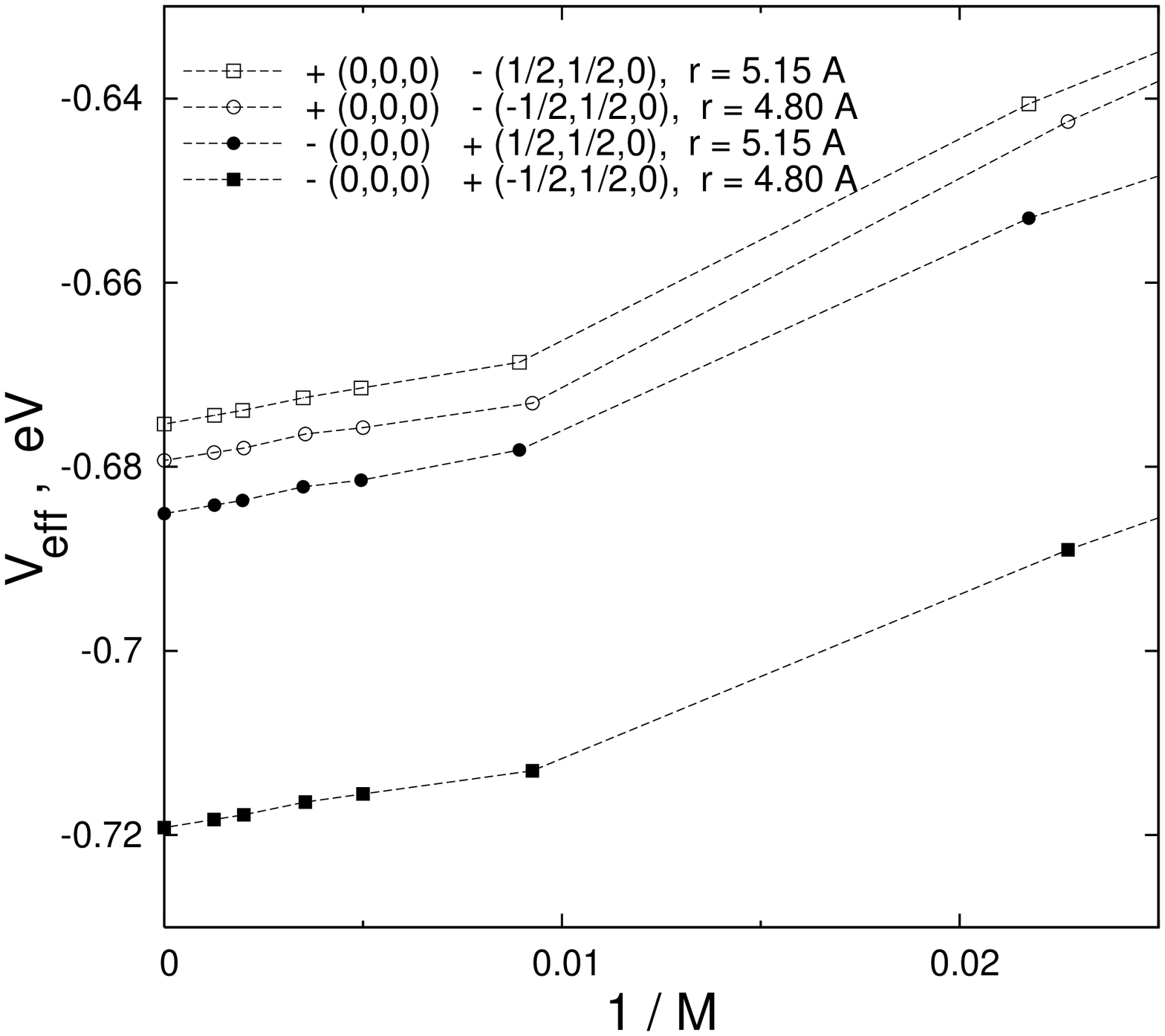,width=2.5 in,angle=-90}}
 \vskip 0.1 in
 {\small {\bf Fig.~3} Convergence of the effective electron-hole
interaction energy with cluster size $M$.}

 \vskip 0.1 in
 {\small {\bf Table II.} Effective interaction energies $V_{\rm
eff}({\bf r})$, in eV, of cation-anion pairs.  The submolecular data
is from \cite{petelenz_slawik_1996}.  Center-to-center distances $r$
are also listed for reference.}

 \centerline{
\begin{tabular}{llrrr}
\\
\tableline
\tableline
 Anion & Cation & $r$, \AA\ & \ \ \ \ \ $V_{\rm eff}({\bf r})$ &
 $ \ \ \ V_{\rm eff}^{\rm submol}({\bf r})$ \\
\tableline
 $(0,0,0)$       & $(-\frac{1}{2},\frac{1}{2},0)$ \ \ \ \ &  $4.7990$ & $-0.719$ & $-0.756$ \\ 
 $(-\frac{1}{2},\frac{1}{2},0)$  & $(0,0,0)$              &  $4.7990$ & $-0.679$ & $-0.756$ \\ 
 $(0,0,0)$       & $(\frac{1}{2},\frac{1}{2},0)$  \ \ \ \ &  $5.1514$ & $-0.685$ & $-0.698$ \\ 
 $(\frac{1}{2},\frac{1}{2},0)$   & $(0,0,0)$              &  $5.1514$ & $-0.675$ & $-0.698$ \\ 
\tableline
\\
\end{tabular}}

It is interesting that the mutual orientation of the cation and ion
together with variations in their electrostatic response are enough to
reverse the order of states, such that the pair with larger
center-to-center distance has greater binding energy (lines 2 and 3 in
Table II).  This is an effect very similar in nature to the one
suggested by Mazur and Petelenz \cite{petelenz_CPL}.

\section{Polarization in Pentacene Films}

\subsection{Inert metal-organic interfaces}

We now address polarization energy in thin pentacene films deposited
on metal surface (Fig.~1).  We use an approach similar to the one
sketched recently for thin films of PTCDA \cite{ours_surface}, which
we present in greater detail.  We assume an ideal metal-organic
interface without chemical interactions.  Organic molecules are
physisorbed and in van der Waals (vdW) contact with the metal.  We
model the metal as an equipotential surface parallel to the organic
layer.  The potential at the metal is assumed to be $\phi=0$, as the
actual value drops out from the combined polarization energy for the
electron and hole, $P=P_++P_-$, even though we perform separate
calculations for $P_+$ and $P_-$.  In fact, any additive constant in
the potential cancels out for any neutral entity with an arbitrary but
equal numbers of positive and negative charges.

For the organic film we use the bulk crystal geometry, assuming a
parallel slab of the crystal cut along molecular layers and placed
parallel to the metal surface.  The only parameter in the model that
allows limited adjustment is the actual distance of the equipotential
plane from the slab.  We choose it such that the equipotential surface
is at the vdW contact distance of 2.80 \AA\ between the closest
hydrogen and gold atoms.  The molecular arrangement depicted in Fig.~1
is in fact drawn to scale.

As in \cite{ours_surface} the equipotential metallic surface is
treated by means of the image charges.  For each atomic partial charge
and induced dipole we assign and place an opposite charge and dipole
in mirror positions when computing potentials and potential gradients.
The charges and dipoles within a molecule do not contribute to the
potentials and potential gradients that act on that molecule.
However, the images of these charges and dipoles do contribute.  For
example, a single ion exhibits no polarization energy in gas phase,
while it acquires polarization energy when placed near the metal
surface.

As in bulk calculations, we solve the polarization problem for slabs
in two steps.  We first consider a neutral film of $N$ molecular
layers with no ions and with translational symmetry in two dimensions.
We solve the self-consistent equations and find ground-state charges
$\overline\rho_i^{kj}$ and dipoles $\overline{\bbox{\mu}}_i^{kj}$,
which are different for each molecular layer $k=1,...,N$ and for each
type $j=1,2$ of molecule in the unit cell.  While the slab is infinite
in two dimensions, the number of variables is finite due to
translational symmetry.

We then replace one molecule in the surface molecular layer by an ion
and solve for the difference
$\delta\rho_i^a=\rho_i^a-\overline\rho_i^{kj}$ and
$\delta\bbox{\mu}_i^a=\bbox{\mu}_i^a-\overline{\bbox{\mu}}_i^{kj}$ for
every molecule $a$ in the slab.  Since the slab is infinite, we define
clusters whose shapes are pillboxes with variable radius $R$ parallel
to the metal and fixed thickness to include $N$ molecular layers and
their images.  We then relax the charge distribution of all $M$
molecules whose centers are in the pillbox.  We assume
$\delta\rho_i^a=\delta\bbox{\mu}_i^a=0$ for the molecules $a$ outside
the pillbox, which means that their charge distributions are not
relaxed.  They contain, however, partial charges and dipoles
$\overline\rho_i^{kj}$ and $\overline{\bbox{\mu}}_i^{kj}$, which
contribute to the total polarization energy of the film.

\subsection{Surface and interface polarization}

The polarization energy for the ion appears, as in the bulk, as a
finite difference of two infinite energies: for the infinite slab with
and without the ion.  The expression for the energy that takes care of
this cancellation is the same as Eq.~(27) in \cite{ours_pplus}, but
with the potentials and potential gradients containing contributions
from image charges and dipoles.  We perform calculations for finite
pillbox clusters of increasing $M$, and extrapolate to
$M\rightarrow\infty$.

Fig.~4 shows the convergence of polarization energies for cations
(upper panel) and anions (lower panel), as well as the extrapolated
values for pentacene films one to ten molecular layers thick.  Two
lines for each film thickness correspond to two types of molecules in
the unit cell.  Convergence with increasing pillbox radius $R$ is
shown as $1/M_l$, where $M_l=M/N$ is the average number of pentacenes
per layer.  The computational effort increases with $N$.  The largest
$N=10$ clusters in Fig.~3 contain $M=2806$ molecules, each containing
36 atoms, and their images.

The combined extrapolated values $P^{\rm S}=P_+^{\rm S}+P_-^{\rm S}$
are also listed in Table III and plotted vs. $1/N$ in Fig.~5.  The
monolayer ($N=1$) and bulk values are almost the same.  Image charges
and dipoles at vdW separation are sufficiently more polarizable to
compensate for a vacuum on the other side.  $P^{\rm S}(N)$ decreases
with increasing $N$ as less polarizable organic layers are introduced
between the surface and the metal.  In the limit of a thick film, or a
free pentacene $ab$ surface, we estimate $P^{\rm S}-P\sim-0.23$ eV.

 \vskip 0.1 in
 {\small {\bf Table III.} Variation in polarization energy at the
surface ($P^{\rm S}$) and at the metal/organic interface ($P^{\rm M}$)
in an N-layer thick pentacene film on a metallic substrate.  The
values are reported for anion at position (1) and cation at position
(2) in the unit cell, which correspond to the lowest energy state.}

 \centerline{
\begin{tabular}{lrr}
\\
\tableline
\tableline
 Layers, $N$ &
\ \ \ \ $(P^{\rm S}-P)$, meV &
\ \ \ \ $(P^{\rm M}-P)$, meV \\
\tableline
  1 &    $6$ &    $6$ \\
  2 &  $-40$ &  $113$ \\
  3 &  $-71$ &  $125$ \\
  5 & $-108$ &  $129$ \\
 10 & $-166$ &  $130$ \\
 thick film & $-227$ &  $130$ \\
\tableline
\\
\end{tabular}}

In Table III the interface values $P^{\rm M}=P_+^{\rm M}+P_-^{\rm M}$
refer to the polarization energy of an ion next to the metal in a film
of $N$ molecular layers.  The biggest increase of 100 meV at $N=2$
comes from the first pentacene overlayer.  Additional layers produce
small changes and the interface value in a thick films is $P^{\rm
M}-P=0.13$ eV.  The hole and electron components of $P^{\rm M}$ are
relevant for matching transport levels to Fermi energies for facile
injection of carriers.  Such optimization for organic devices is
largely empirical at present, often without {\em any} consideration of
polarization.  The variation of $P^{\rm S}(N)$ or $P^{\rm M}(N)$ with
film thickness is less in pentacene than in PTCDA \cite{ours_surface}.
Pentacene layers are several times thicker, with the long axis normal
to the metal, and a single layer already provides an effective
polarization environment for the charge.

 \vskip 0.1 in
 \centerline{\epsfig{file=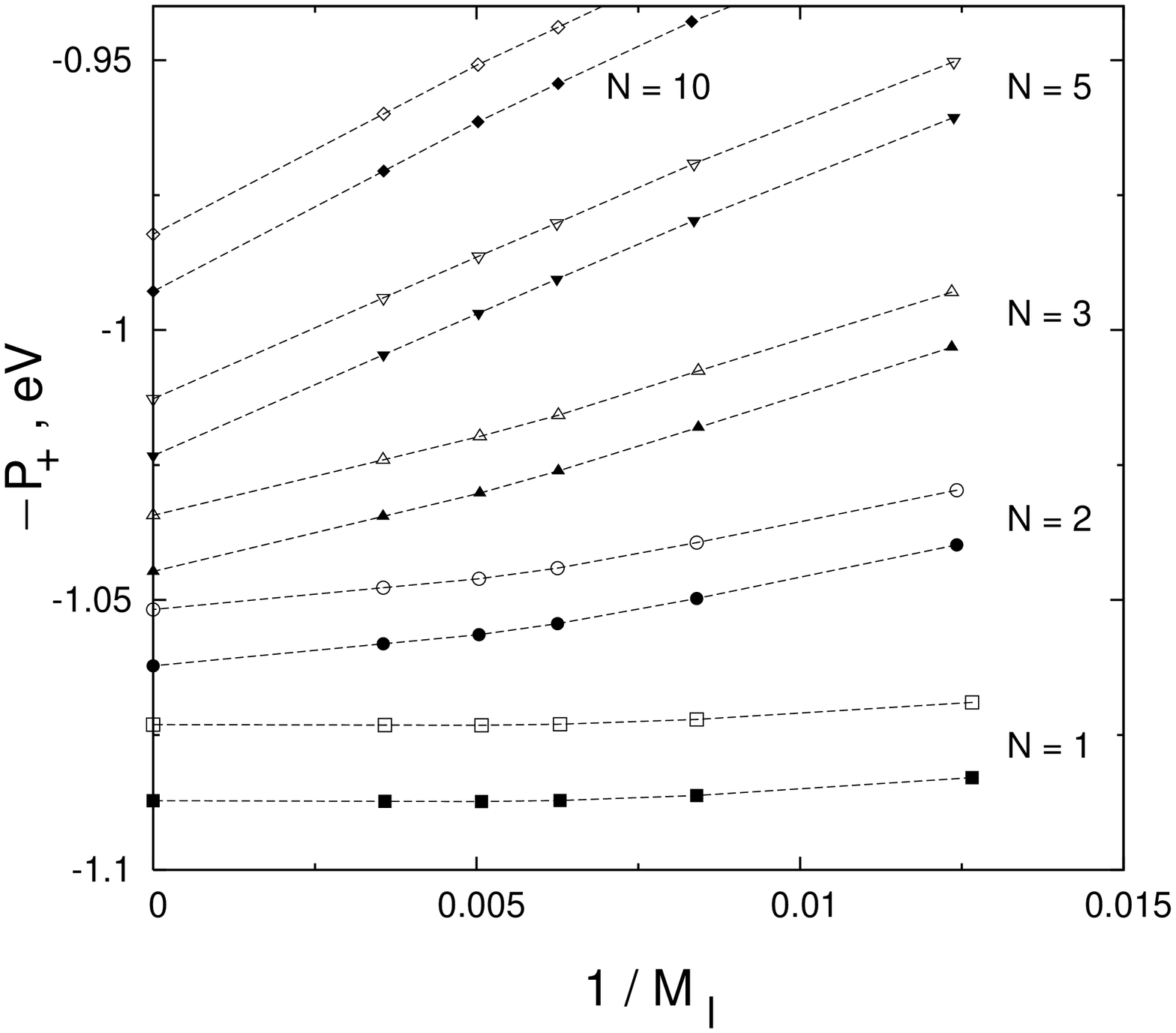,width=2.5 in,angle=-90}}
 \centerline{\epsfig{file=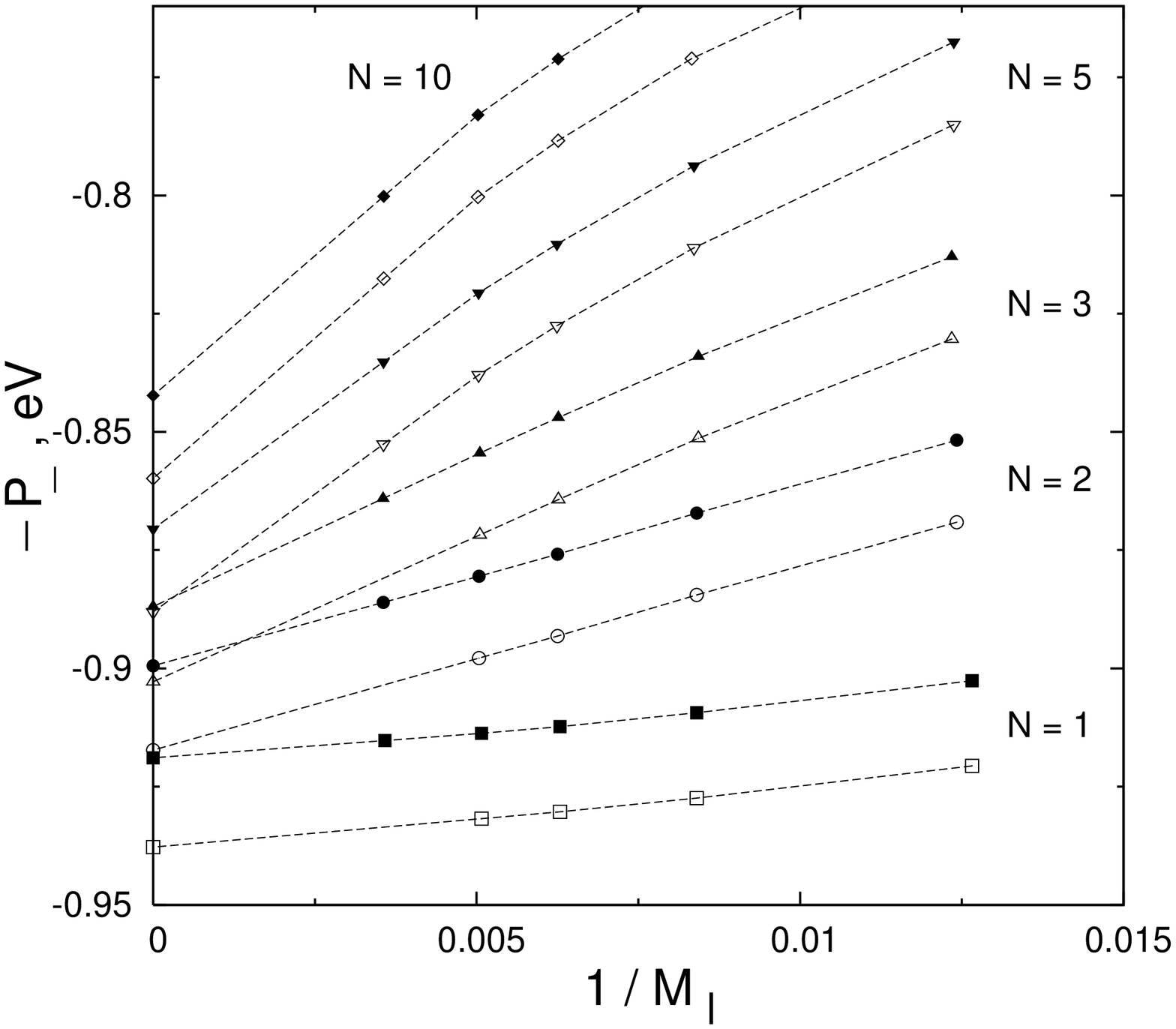,width=2.5 in,angle=-90}}
 \vskip 0.1 in
 {\small {\bf Fig.~4} Convergence of polarization energies of a cation
(upper panel) and anion (lower panel) in the outermost surface layer
of pentacene film on gold vs. the average number $M_l=M/N$ of
molecules in each molecular layer ($M$ molecules in $N$ layers in the
pillbox).  Open and solid symbols correspond to the molecules of type
(1) and (2) respectively.}
 \vskip 0.1 in

 \vskip 0.1 in
 \centerline{\epsfig{file=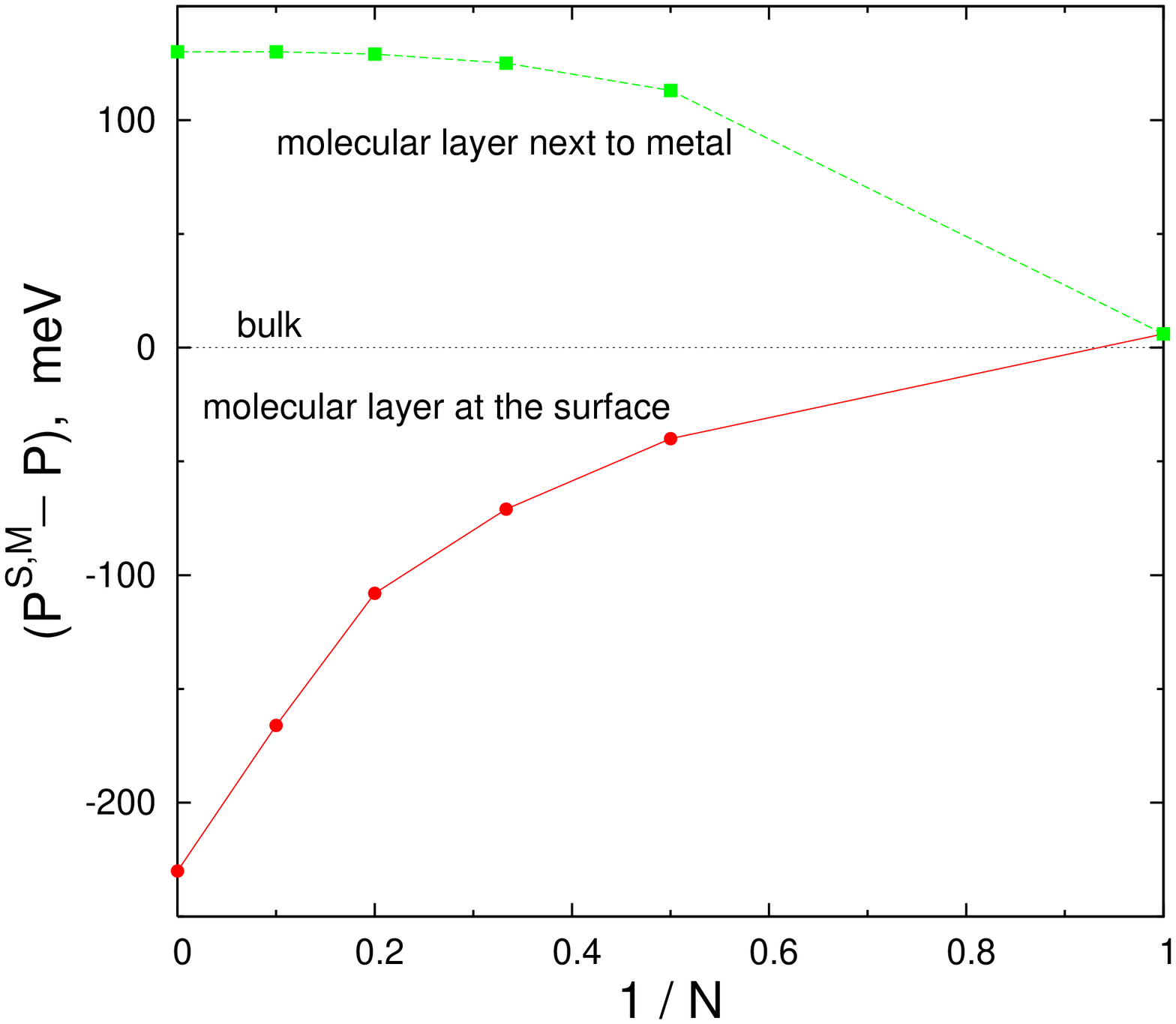,width=2.5 in,angle=-90}}
 \vskip 0.1 in
 {\small {\bf Fig.~5} Variation in polarization energy at the surface
($P^{\rm S}(N)-P$) and at the metal/organic interface ($P^{\rm
M}(N)-P$) for an $N$-layer thick pentacene film on a metal substrate.}

\subsection{Molecular potential}

We have already noted the competition between image charges and
induced dipoles in films of molecules with large polarizability normal
to the surface.  The self-consistent potential $\phi^a({\bf r})$ at
molecule $a$ is due to the polarized densities $\rho^b({\bf r})$ of
all other molecules and their images.  A single pentacene cation in
the surface layer induces an image anion in the metal.  Since the
interaction is attractive, the cation's $\phi^{cat}({\bf r})$ increase
monotonically with the distance from the surface.  Similarly, the
$\phi^{an}({\bf r})$ of a single anion on the surface decrease
monotonically with distance from the surface.

The situation in films is quite different due to the surrounding
neutral molecules.  Since the large pentacene polarizability is normal
to the surface, the induced moments due to the image anion are
parallel.  The repulsion of parallel dipoles and image dipoles is
relieved by redistributing charge towards the middle of the molecular
layer.  Now the $\phi^{cat}({\bf r})$ has a minimum at atoms near the
center, while the $\phi^{an}({\bf r})$ has a maximum near the center.
The actual $\phi^{ion}({\bf r})$ in the limit of zero overlap is the
solutions of the self-consistent equations.

Figure 6 shows the self-consistent potential $\phi^{cat}({\bf r})$ for
${\bf r}$ along the axis of a cation in a pentacene monolayer on
metal.  The gradients yield electric fields $\pm10^7$ V/cm that
reverse over a single molecule, directly confirming that fields are
strong and non-uniform.  The variation of $\phi^{cat}({\bf r})$ with
${\bf r}$ shows the importance of repulsion between induced dipoles.
Image charges account for more negative $\phi^{cat}({\bf r})$ close to
metal.  Similar variations of self-consistent potentials are found at
ions (1) and at neighboring neutral molecules.  The strong shielding
of a single layer of pentacene follows from the weak dependence on
$N$.  Similar curves (not shown) for $N=2$, 3, and 5 are essentially
parallel to the $N=1$ curve in Fig.~6, but displaced to less negative
(positive) potential for cations (anions).

 \vskip 0.1 in
 \centerline{\epsfig{file=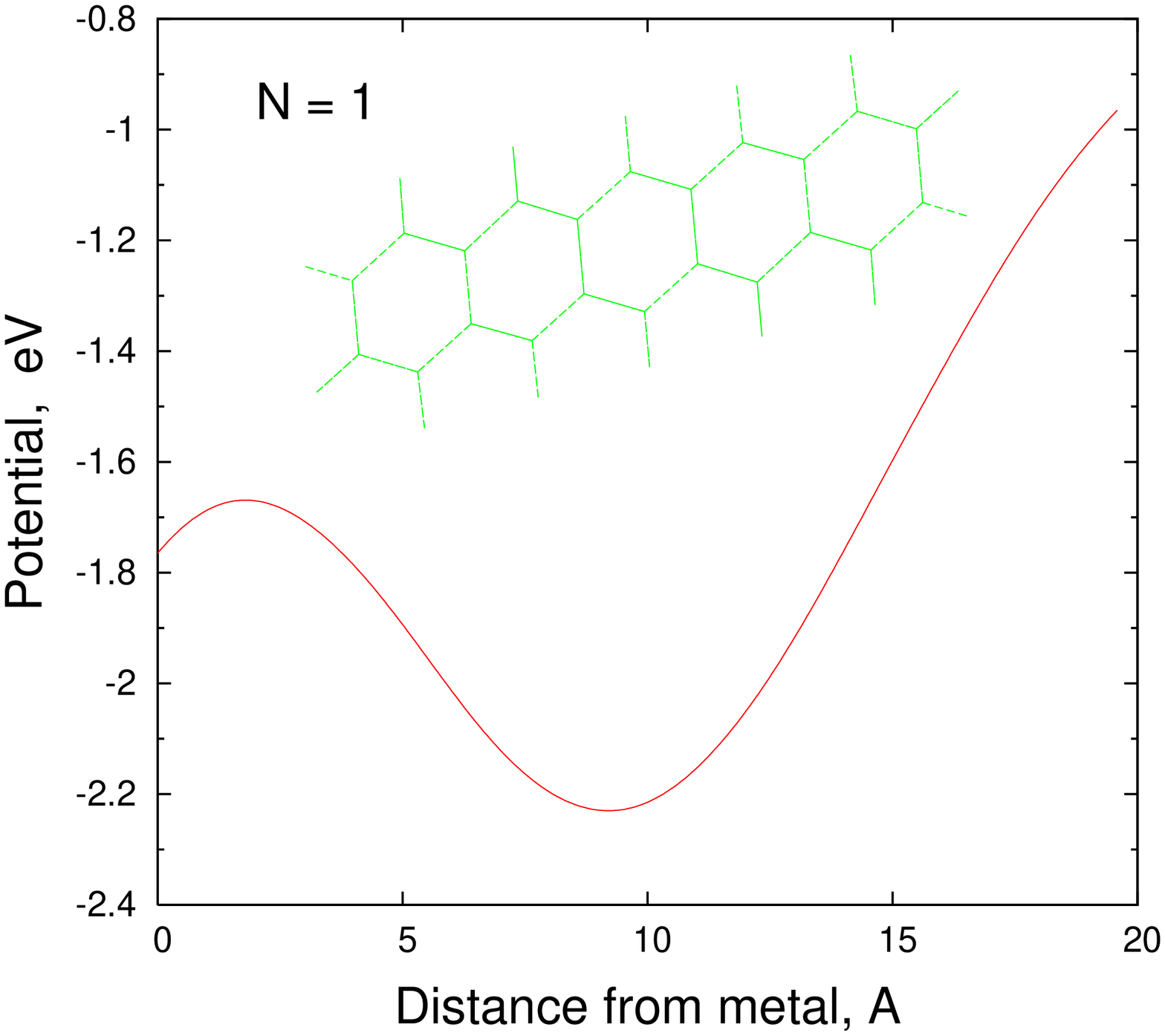,width=2.5 in,angle=-90}}
 \vskip 0.1 in
 {\small {\bf Fig.~6} Self-consistent potential for a cation created
by the polarization field of neutral molecules in the lattice and the
image charge distribution in the metal.  Data shown for a cation in a
monolayer on metal substrate, placed in the position (2) in the unit
cell.  The sketch of the molecule shows actual position with respect
to the metal.}

\section{Discussion}

The bulk polarization energy can be estimated simply in terms of
molecular size, $P_\pm=(e^2/2R)(1-1/\kappa)$, for a sphere of radius
$R$ in an isotropic dielectric medium with $\kappa\sim3$ for organic
molecular crystals.  Such rough values rationalize much data but
preclude accurate positioning of transport or CT states.  The
effective pentacene radius is 14\% greater than that of anthracene,
while the polarization energy decreases by only 9\%.  On the other
hand, the radii of pentacene and PTCDA molecules are the same within
3\%, but the polarization energy differs by 10\%.  The greatest
contrast to PTCDA films \cite{ours_surface}, is that the polarization
energy $P^{\rm S}$ is within 10\% of the bulk value in pentacene,
compared to $P^{\rm S}/P=0.77$ in PTCDA.

Direct relations between gas-phase and crystalline energies have long
been sought using molecular exciton theory \cite{silinsh,pope}.
Accurate evaluation of electronic polarization permits sharper
comparisons.  There is considerable scatter in reported values for
both the gas and solid, as noted in Sections III(A) and III(D) for the
ionization potential and electron affinity of pentacene.  The
transport gap is $I-A-P$ in the limit of zero intermolecular overlap.
The experimental values for the pentacene transport gap range from
2.85 eV \cite{petelenz_slawik_1996} to 2.4 eV suggested in
\cite{sebastian_1981}, to 2.2 eV used in \cite{kahn_2002}.  This
scatter was not ``important'' when polarization energies had to be
estimated in the bulk and were not available at all at surfaces.  With
the ability to calculate polarization energies, it becomes possible to
analyze and connect various sources of experimental data on such
quantities as optical and transport gaps, CT binding energies, and
work functions.

All molecular calculations, except for the gas-phase properties, were
performed using experimental geometries from the crystal structure
data \cite{pentacene_structure}.  Slight variations in the geometry of
the two molecules in the unit cell were consequently taken into
account.  While the variations in electronic energies are small (tens
of meV), which is consistent with the adopted approximation, we
believe that crystalline geometries are preferable to gas-phase
geometries when vertical electronic transitions are considered in the
condensed phase.

We note that INDO/S greatly overestimates the in-plane
polarizabilities of ions in Table I.  The huge difference of
$\alpha^{\rm C}$ between cations 1 and 2 is clearly unphysical in view
of small distortions in the crystal, and it rationalizes why
convergence was particularly difficult for cation 1.  The UHF INDO/S
approach fails for ions.  Radical ions of large conjugated molecules
are expected to be more polarizable than the neutral molecules, in
line with the B3LYP entries in Table I.  While the idea is that
$\widetilde\alpha$ constitutes a small correction, and it is not so
for ions, we expect this to have a negligible impact on the results.
The reason is that the largest part of the polarization energy comes
not from the ion itself but from the polarization of many neutral
molecules surrounding it.  Moreover, the molecular $\alpha$ is kept no
matter how it is partitioned between charge redistribution
($\alpha^{\rm C}$) and induced dipoles.  In the future we may choose
to use $\Pi_{ij}$ of the neutral molecule for the ions, and to absorb
the polarizability difference in $\widetilde\alpha$.  This requires
separate B3LYP calculation of $\alpha$ for the ions, as done here, but
not in \cite{ours_pplus}.

Small atomic charges in acenes are related to approximate electron-hole
symmetry. The crystal potential $\phi_i=\phi({\bf r}_i)$ at atom $i$
due to INDO/S charges is consequently small, as shown by $\sim1$ meV
per molecule contributions to the sublimation energy. Since the
potential due to the B3LYP or any other gas-phase charge distribution
can readily be computed at the positions of other atoms in the
crystal, we can obtain first-order corrections to polarization
energies in general \cite{europhys} by combining the best available
gas-phase potential with self-consistent charges and dipoles based in
INDO/S calculations. The quadrupole moments of acenes lead to opposite
shifts of $P_+$ and $P_-$, as discussed previously for submolecules
\cite{silinsh,munn}.  Since such corrections cancel in $P=P_++P_-$ in
anthracene \cite{ours_pplus}, we expect them to be small in pentacene.
They are not negligible in PTCDA or conjugated molecules that contain
heteroatoms and hence partial charges.

Our values for the the transport gap are likely an overestimate, due
to the bandwidth effects for electrons and holes, and the correction
can be introduced as $E_t=I-A-P-W$.  The bandwidth $W$ in molecular
crystals at room temperature is thought to be of the order of few tens
of meV or less, which justifies our zero-overlap approximation.  At
lower temperatures, however, the bandwidths may be large in
certain directions.  It has been suggested that the bandwidth in
pentacene (at zero temperature) may be anomalously large and reach 500
meV \cite{bredas_2001}.  If we estimate $W\sim0.5$ eV, we get
$E_t=2.71$ eV with the experimental charge gap, while the calculated
charge gap leads to the value 2.35 eV, which is too small.

When bandwidths are comparable to polarization energies, the
zero-overlap approximation cannot be justified, and we face a
challenging problem of combining the ``localized'' treatment of
polarization effects with the ``delocalized'' language of band theory.
The self-consistent polarization approach can provide a proper
``zero-order'' approximation for the attempts to treat bandwidth
effects as perturbation.

We note that our self-consistent approach has the ability to
distinguish between the electronic polarization of cations and anions,
since the quantum-mechanical structure of molecular response is used
to obtain the redistribution of charge.  We can also estimate
variations of polarization caused by slight changes in molecular
geometry between crystal and gas-phase, and especially between
inequivalent molecules in the unit cell.  These variations appear to
be of the order of 20---40 meV, which is on the threshold of the
accuracy, but such variations are probably correct on the order of
magnitude, and may be important in the analysis of experimental data.

In summary, we have applied the recently developed self-consistent
approach based on evaluating charge redistribution in organic
molecules to the problem of electronic polarization in pentacene
molecular crystals.  The power of the approach comes from the
combination of semiempirical treatment of charge distribution by means
of atom-atom polarizability tensors with accurate {\em ab initio}
gas-phase calculations of molecular polarizabilities.  As a result, we
are now able to calculate polarization energies with accuracy on the
order of 0.1 eV or better.

\vskip 0.3 in

\centerline{\bf ACKNOWLEDGMENTS}

\vskip 0.1 in

It is a pleasure to thank A. Kahn and R.A. Pascal, Jr. for stimulating
discussions.  We gratefully acknowledge support for work at Rutgers
through the Office of Naval Research, grant No.~NO00014-01-1-1061, and
at Princeton through the MRSEC program under DMR-9400632.

\end{multicols}
\end{document}